\documentclass[floatfix,aps,rmp,groupedaddress]{revtex4}

\usepackage[T2A]{fontenc}
\usepackage[utf8x]{inputenc} 
\usepackage[american]{babel}

\usepackage{epsfig}
\usepackage{amsmath}
\usepackage{amssymb}
\usepackage{array}

\newcommand{\crit}[1]{\ensuremath{{#1}^{\star}}}

\newcommand{\sat}[1]{{#1}_\mathnormal{\sigma}}

\newcommand{\qq}[1]{\lq{#1}\rq}

\newcommand{\grandpotentialof}[1]{\ensuremath{{\grandpotential}_{#1}}}
\newcommand{\nuclrateof}[1]{\ensuremath{{\nuclrate}_{#1}}}

\newcommand{\probinftyof}[1]{\ensuremath{\probinfty_{#1}}}

\newcommand{\absnum}{\ensuremath{\mathnormal{N}}}
\newcommand{\area}{\ensuremath{\mathcal{F}}}
\newcommand{\chempot}{\ensuremath{\mathnormal{\mu}}}

\newcommand{\committor}{\ensuremath{\mathnormal{q}}}

\newcommand{\density}{\ensuremath{\mathnormal{\rho}}}
\newcommand{\differential}{\ensuremath{\mathnormal{d}}}

\newcommand{\grandpotential}{\ensuremath{\mathnormal{\Omega}}}
\newcommand{\deltahvap}{\ensuremath{\Delta\mathnormal{h}^\mathrm{v}}}

\newcommand{\kboltz}{\ensuremath{\mathnormal{k}_\mathrm{B}}}

\newcommand{\LJeps}{\ensuremath{\mathnormal{\varepsilon}}}
\newcommand{\LJsig}{\ensuremath{\mathnormal{\sigma}}}
\newcommand{\mass}{\ensuremath{\mathnormal{m}}}

\newcommand{\nucdensity}{\ensuremath{\density_\mathrm{n}}}
\newcommand{\nuclsize}{\ensuremath{\mathnormal{\nu}}}
\newcommand{\nuclrate}{\ensuremath{\mathcal{J}}}
\newcommand{\nuclrateCNT}{\ensuremath{\nuclrate_\mathrm{CNT}}}
\newcommand{\nuclrateHale}{\ensuremath{\nuclrate_\mathrm{HSL}}}

\newcommand{\nucradius}{\ensuremath{\mathcal{R}}}
\newcommand{\NVT}{\ensuremath{NVT}}
\newcommand{\overheating}{\ensuremath{\Delta\temperature}}
\newcommand{\partition}{\ensuremath{\mathcal{Z}}}
\newcommand{\planartension}{\tension_\infty}

\newcommand{\preexp}{\ensuremath{\mathnormal{A}}}
\newcommand{\pressure}{\ensuremath{\mathnormal{p}}}
\newcommand{\probability}{\ensuremath{\mathrm{P}}}
\newcommand{\probgrow}{\ensuremath{\mathnormal{w}}}
\newcommand{\probinfty}{\committor}

\newcommand{\steric}{\ensuremath{\mathnormal{s}}}
\newcommand{\supersat}{\ensuremath{\mathnormal{S}}}

\newcommand{\temperature}{\ensuremath{\mathnormal{T}}}
\newcommand{\Tc}{\ensuremath{\temperature_\mathrm{c}}}
\newcommand{\Ttr}{\ensuremath{\temperature_3}}
\newcommand{\tension}{\ensuremath{\mathnormal{\gamma}}}
\newcommand{\thermallength}{\ensuremath{\mathnormal{\Lambda}}}
\newcommand{\threshold}{\ensuremath{\mathnormal{\Theta}}}
\newcommand{\timea}{\ensuremath{\mathnormal{t}}}
\newcommand{\tolmanlength}{\ensuremath{\mathnormal{\delta}}}
\newcommand{\upot}{\ensuremath{\mathcal{V}}}
\newcommand{\Deltaupot}{\ensuremath{\delta\upot}}
\newcommand{\volume}{\ensuremath{\mathnormal{V}}}
\newcommand{\zeldovich}{\ensuremath{\mathnormal{f}_\mathrm{Z}}}

\newcommand{\daemon}{demon}
\newcommand{\LJTS}{tsLJ}
\newcommand{\Nucleus}{Nucleus}
\newcommand{\nucleus}{nucleus}
\newcommand{\Nuclei}{Nuclei}
\newcommand{\nuclei}{nuclei}

\newcommand{\vapour}{vapor}
\newcommand{\vapours}{vapors}

\begin{document}
\begin{abstract}
   Grand canonical molecular dynamics (GCMD) is applied to
   the nucleation process in a metastable phase near 
   the spinodal, where nucleation occurs almost instantaneously
   and is limited to a very short time interval.
   With a variant of Maxwell's demon, proposed by
   McDonald [Am.\ J.\ Phys.\ 31 (1963): 31], all nuclei exceeding a
   specified size are removed.
   In such a steady-state simulation, the nucleation process
   is sampled over an arbitrary timespan
   and all properties of the metastable state,
   including the nucleation rate, can be obtained with an increased
   precision.
   As an example, a series of GCMD
   simulations with McDonald's demon is carried out for 
   homogeneous vapor to liquid nucleation of the truncated-shifted
   Lennard-Jones (tsLJ) fluid, covering the entire relevant temperature range.
   The results are in agreement with direct non-equilibrium
   MD simulation in the canonical ensemble.
   It is confirmed for supersaturated vapors
   of the tsLJ fluid that the classical nucleation theory underpredicts the
   nucleation rate by two orders of magnitude.
\end{abstract}

\title{Grand canonical steady-state simulation of nucleation}

\date{\today}

\author{Martin Horsch}
\author{Jadran Vrabec\footnote{Corresponding author. Email: jadran.vrabec@upb.de}}
\affiliation{Universit\"at Paderborn, Lehrstuhl f\"ur Thermodynamik und Energietechnik,
                                      Warburger Str.\ 100, 33098 Paderborn, Germany}
\maketitle

\section{Introduction}
The key properties of nucleation processes are the height
$\crit{\Delta\grandpotential}$ of the free energy barrier that must be
overcome to form stable embryos of the emerging phase and the
nucleation rate $\nuclrate$ that indicates how many \nuclei{}
appear in a given volume per time.
The most widespread approach for calculating these quantities is
the classical nucleation theory \cite{FRLP66}, which has significant
shortcomings, e.g., it can overestimate
$\crit{\Delta\grandpotential}$ significantly for homogeneous
\vapour{} to liquid nucleation \cite{Talanquer07}.
A more accurate theory of homogeneous nucleation, which is sought after,
would also increase the reliability for more complex applications such as
heterogeneous and ion-induced nucleation in the earth's atmosphere.

An important problem of the classical nucleation theory (CNT) is that the underlying basic assumptions
do not apply to nanoscopic \nuclei{} \cite{VHH09}.
Although it is possible to measure the critical size by
neutron scattering \cite{Debenedetti06, PRCB06},
the thermophysical properties of such \nuclei{} are
mostly very hard to investigate experimentally.
However, they are well accessible by calculations based on
density functional theory \cite{OE88, ZO91, BS99, UC08}
as well as molecular simulation \cite{VKFH06, HL08, SVB09}.
For instance, vaporization processes \cite{HL08}
and equilibria \cite{VKFH06, SVB09}
of single liquid droplets can be simulated to obtain the surface tension as
well as heat and mass transfer properties of strongly curved interfaces.
Similarly, very fast nucleation processes that occur in the immediate vicinity of
the spinodal are experimentally inaccessible, whereas they can be studied by
Monte Carlo \cite{NV05} and molecular dynamics \cite{YM98, HVH08}
simulation of systems with a large number of particles.
Lower nucleation rates are accessible by transition path sampling based
methods such as forward flux sampling \cite{GVLMFF08, MPSF08}.
Hence, molecular simulation is crucial for the further
development of nucleation theory.

Such molecular dynamics (MD) simulations, dealing with single \nuclei{} in equilibrium as well as
with homogeneous nucleation processes in supersaturated \vapours{}, led to the formulation
of a surface propery corrected (SPC) modification of CNT for \vapour{} to liquid
nucleation of unpolar fluids, cf.\ previous
work \cite{HVH08} for a detailed presentation
and justification. Both CNT and the SPC modification apply the expression
\begin{equation}
   \left(\frac{\partial\grandpotential}{\partial\nuclsize}\right)_{\chempot, \volume, \temperature}
      = \tension\left(\frac{\partial\area}{\partial\nuclsize}\right)_{\chempot, \volume, \temperature}
           - \left[\chempot - \sat\chempot(\temperature)\right],
\end{equation}
accounting for the positive contribution of the surface tension
$\tension$ acting on the surface area $\area$ as well as a negative
bulk contribution, where $\chempot$ and $\sat\chempot(\temperature)$
are the chemical potential of the supersaturated and the
saturated \vapour{} at the temperature $\temperature$,
respectively, to the free energy of formation
\begin{equation}
   \Delta\grandpotentialof{\nuclsize} = \int_{1}^{\nuclsize}
      \left(\frac{\partial\grandpotential}{\partial\nuclsize}\right)_{\chempot, \volume, \temperature}
         \differential\nuclsize,
\end{equation}
for a \nucleus{} containing $\nuclsize$ particles. The maximal 
free energy of formation $\crit{\Delta\grandpotential}$, corresponding
to the critical size $\crit\nuclsize$, is the decisive
quantity for the nucleation rate, given by the Arrhenius equation
as
\begin{equation}
   \nuclrate = \preexp\exp\left(\frac{- \crit{\Delta\grandpotential}}{\kboltz\temperature}\right),
   \label{eqn:J}
\end{equation}
where $\kboltz$ is the Boltzmann constant. In addition to the usual
collision term from kinetic gas theory, the pre-exponential coefficient $\preexp$
includes the Z\v{e}l'dovi\v{c} (Зельдович)
factor and a correction for thermal non-accomodation \cite{FRLP66}.
Although $\preexp$ is not constant, it depends to a much lower extent
on supersaturation than the exponential term. CNT applies the
capillarity approximation, which in the present context means that $\tension$
is assumed to be the same as the surface tension of the planar \vapour-liquid
interface $\planartension$. The surface area is determined from the
assumption that all \nuclei{} are spherical. The SPC modification
replaces the capillarity approximation with the Tolman equation \cite{Tolman49},
\begin{equation}
   \frac{\planartension}{\tension} = 1 + \frac{2\tolmanlength}{\nucradius},
\end{equation}
wherein $\nucradius$ is the radius of the \nucleus{} and $\tolmanlength$
is the Tolman length, a characteristic interface thickness, while the surface
area is increased by a steric factor $\steric$. In particular, the
temperature-dependent correlations
\begin{equation}
   \tolmanlength\slash\nucradius
      = \left(\frac{0.7}{1 - \temperature\slash\Tc} - 0.9\right)
         \nuclsize^{-1\slash{}3},
\end{equation}
with respect to the critical temperature $\Tc$, as well as
\begin{equation}
   \steric =
      \frac{0.85 \, (1 - \temperature\slash\Tc)^{-1} + (\nuclsize\slash{}75)^{1\slash{}3}
         }{1 + (\nuclsize\slash{}75)^{1\slash{}3}},
\end{equation}
can be used for unpolar fluids \cite{HVH08}.
A different approach is given by the Hale scaling law (HSL). 
In agreement with experimental data on nucleation of water and toluene \cite{Hale86}, it predicts
\begin{equation}
   \nuclrate \sim \density^{-2\slash{}3}\left(\frac{\planartension}{\temperature}\right)^{1\slash{}2}\pressure^2
      \exp\left( \frac{4(\kboltz\temperature)^2\planartension^3}{27 (\chempot - \sat\chempot)^2} \right),
\end{equation}
where the proportionality constant only depends on properties of the critical point.

The present work has the
objective of refining the methodology used for direct MD simulation of nucleation processes.
According to the method of Yasuoka and Matsumoto (YM),
a supersaturated \vapour{} is simulated in the canonical ensemble and the nucleation rate is obtained
from the number of \nuclei{} formed over time, using a linear fit
where only \nuclei{} that exceed a sufficiently large threshold size are counted \cite{YM98}.
Nucleation occurs after the metastable state is equilibrated and before \nucleus{}
growth becomes dominant.
However, the timespan corresponding
to nucleation is very short for the high nucleation rates that
are accessible to direct MD simulation, which restricts the statistical
basis and the precision of the results. Near the spinodal,
the regimes of equilibration, nucleation, and growth even start to overlap
and the YM method becomes unreliable.

Wedekind \textit{et al.}\ recently developed a more rigorous method which
is based on mean first passage times (MFPT) obtained by averaging over hundreds
of simulation runs \cite{WSR07}.
But as Chkonia \textit{et al.}\ point out, \qq{the computational costs of
making the necessary repetitions to evaluate the MFPT can be very high,} whereas
\qq{YM requires many clusters forming and it therefore becomes more sensible to
deviations coming from vapor depletion or coalescence of clusters} \cite{CWSWR09}.

A new direct equilibrium MD simulation method is introduced in the present work.
The underlying concept is to simulate the non-equilibrium as a stationary
process in the grand canonical ensemble. Thereby, it is possible to sample exclusively nucleation
as opposed to \nucleus{} growth and coalescence.
While the precision of the results
is increased by maintaining the steady state over an arbitrarily long
time interval, the advantages of the YM non-equilibrium method are also retained.
In particular, only one MD simulation run
is required and the nucleation rate is obtained from the number
of large \nuclei{} formed over time.
This is achieved by combining grand canonical molecular dynamics (GCMD), introduced by
\cite{Cielinski85}, and an \qq{intelligent being} that 
continuously removes all large \nuclei{} \cite{McDonald62}:
McDonald's \daemon.

\section{Simulation method}
%
%
In a closed system, nucleation is an instationary process because
the metastable phase is depleted by the emerging \nuclei{}.
The idea behind the
present approach is to simulate the production of \nuclei{} up to a given size for
a specified metastable state. \Nuclei{} above the given
size are extracted, and particles are inserted as monomers into the system
to replenish the metastable phase.

GCMD regulates the chemical potential
and samples the grand canonical ensemble: alternating
with standard MD steps, particles are deleted from and inserted into
the system probabilistically with the usual grand canonical
acceptance criteria \cite{Cielinski85, LS91}.
For a test deletion, a random particle is removed.
For a test insertion, the coordinates of an additional particle
are chosen at random.
The potential energy difference $\Deltaupot$ is determined for each of the test operations
and compared with the residual chemical potential.
%
%
%
The acceptance probability is defined the same way as for the
Metropolis algorithm, i.e., it is
\begin{equation}
   \probability = \min\left(
      \density\thermallength^3
         \exp\left[\frac{- \chempot - \Deltaupot}{\kboltz\temperature}\right],
            1\right),
\end{equation}
in case of deletions and similar for insertions \cite{AT87}.
In this expression,
$\density$ is the density and $\thermallength$ is the thermal wavelength.
The number of test deletions and insertions per simulation time step
was chosen in this work between $10^{-6}$ and $10^{-3}$ times the number of particles.

Whenever a \nucleus{} exceeds the specified
threshold size $\threshold$, McDonald's \daemon{} \cite{McDonald62} --
called Szil\'ard's \daemon{} by \cite{SRL97} --
removes it from the system and replaces it by a representative
configuration of the metastable phase.
If a dense phase is simulated, this can be achieved by, e.g., inserting
an equilibrated homogeneous configuration in the center of the free volume,
followed by preferential test insertions and deletions in the affected region.
In a supersaturated \vapour{}, however, the density
is usually so low that it is sufficient to leave a vacuum behind
as suggested by \cite{McDonald62}.

Establishing a steady state by continuously removing the largest \nuclei{}
is the purpose and the main advantage of McDonald's \daemon{}. Consequently,
the further behavior of these \nuclei{} cannot be tracked. It is assumed
that most of the \nuclei{} that are extracted would have continued to grow
and that the \daemon{} intervention rate
$\nuclrateof{\threshold}$ is therefore similar to the actual nucleation rate $\nuclrate$.
The deviation between these rates can also be quantified by regarding the size
evolution of a single \nucleus{} in terms of a discrete one-dimensional random
walk over the order parameter $\nuclsize$. At each size transition, $\nuclsize$
is either decreased or increased by one \cite{HMV09}. The short-term growth
probability, corresponding to a size increase in the next step, is then given by
\begin{equation}
   \probgrow = \left(1 + \frac{\partition_{\nuclsize - 1}}{\partition_{\nuclsize + 1}}\right)^{-1},
\end{equation}
where $\partition_{\nuclsize\pm{}1}$ is the grand canonical partition function under the
condition that the \nucleus{} contains $\nuclsize\pm{}1$ particles.
Neglecting all discrete size effects, the long-term growth probability $\probinftyof{\nuclsize}$,
which corresponds to the cases where the \nucleus{} never evaporates completely and thus
eventually reaches arbitrarily large sizes, has the property \cite{HMV09}
%
%
\begin{equation}
   \frac{\differential\partition}{\partition\differential\nuclsize}
      = \frac{-\differential\left(\differential\probinftyof{\nuclsize}\slash\differential\nuclsize\right)
         }{2\left(\differential\probinftyof{\nuclsize}\slash\differential\nuclsize\right)
	    \differential\nuclsize}.
\end{equation}
Using adequate boundary conditions, the long-term growth probability
of the \nuclei{} that are removed by the \daemon{} can be determined as
\begin{equation}
   \probinftyof{\threshold}
      = \frac{\int_1^\threshold
         \partition_\nuclsize^{-2} \differential\nuclsize
            }{ \int_1^\infty
	       \partition_\nuclsize^{-2} \differential\nuclsize
	       }.
   \label{eqn:committor}
\end{equation}
The intervention rate is therefore related to the nucleation
rate by
\begin{equation}
   \nuclrateof{\threshold}
      \int_1^\threshold \exp\left(
         \frac{2\Delta\grandpotentialof{\nuclsize}}{\kboltz\temperature}
	    \right) \differential\nuclsize 
	       = \nuclrate \int_1^\infty \exp\left(
                  \frac{2\Delta\grandpotentialof{\nuclsize}}{\kboltz\temperature}
	             \right) \differential\nuclsize.
\label{eqn:intervention}
\end{equation}
In particular, for a threshold size
sufficiently above $\crit{\nuclsize}$ the approximation
$\nuclrate \approx \nuclrateof{\threshold}$
is valid \cite{HMV09}.


The truncated-shifted Lennard-Jones (\LJTS{}) fluid
accurately describes the fluid phase coexistence of noble gases
and methane \cite{VKFH06}, avoiding long-range corrections which are
tedious for inhomogeneous systems.
Homogeneous \vapour{} to liquid
nucleation of the \LJTS{} fluid was studied here by GCMD simulation
with McDonald's \daemon{} at temperatures of 0.65 to 0.95 in units
of $\LJeps\slash\kboltz$ (where $\LJeps$ is the energy parameter of
the Lennard-Jones potential).
Note that the triple point temperature of the \LJTS{} fluid is
$\Ttr$ = 0.65 while $\Tc$ is 1.078 so that the entire relevant temperature
range is covered \cite{VKFH06, MPSF08}.
The \cite{Stillinger63} criterion was used to discern
the emerging liquid from the surrounding supersaturated \vapour{}
and \nuclei{} were determined as biconnected components. 

\section{Simulation results}
Figure \ref{szilII} shows the aggregated number of \daemon{} interventions
in one of the present GCMD simulations and, for comparison, the number of \nuclei{}
in a MD simulation of the canonical ensemble under similar conditions.
The constant value of the supersaturation
\begin{equation}
   \supersat = \exp\left(\frac{\chempot - \sat\chempot(\temperature)}{\kboltz\temperature}\right),
\end{equation}
in the GCMD simulation agreed approximately with the
time-dependent $\supersat$ in the $\NVT$ simulation about
$\timea$ = 400 after simulation onset in units
of $\LJsig(\mass\slash\LJeps)^{1/2}$, wherein
$\LJsig$ is the size parameter of the Lennard-Jones potential and
$\mass$ is the mass of a particle.

During the $\NVT$ run, however, $\supersat$
decreased from about 3 to 1.5. The observed rate of formation was significantly lower for
larger nuclei, which is partly due to the the depletion of the \vapour{} over simulation
time. Depletion causes less monomers to interact with a \nucleus{}
surface when large \nuclei{} are
formed because by that time, a substantial amount of particles already belong to the liquid.
Moreover, a small nucleus will eventually decay with a
higher probability, given by $1 - \committor$,
instead of growing to arbitrarily large sizes, cf.\ Eq.\ (\ref{eqn:committor}).
Therefore, large \nuclei{} are necessarily formed at a lower rate.

In Fig.\ \ref{szilVI}, it can be seen how the decreasing supersaturation in the
canonical ensemble MD simulation affects the \nucleus{} size distribution.
Around $\timea$ = 400, the distribution of
small \nuclei{} present per volume was similar in both
simulation approaches. Near and above
the critical size, i.e., 27 particles according to CNT, cf.\ Tab.\ \ref{tabJ},
deviations arise because of the different boundary conditions.
Comparing the distribution for the grand canonical steady state
with the corresponding theoretical prediction shows that CNT
underestimates the number of \nuclei{} present in the metastable state,
confirming the result of \cite{Talanquer07} that CNT exaggerates
the free energy of \nucleus{} formation.

CNT is also known to underestimate the nucleation rate of unpolar
fluids \cite{HVH08}.
The determined \daemon{} intervention rates confirm this conclusion,
cf.\ Tab.\ \ref{tabJ},
and as shown in Fig.\ \ref{szilIV},
the HSL is significantly more accurate than CNT 
for low temperatures.
For $\temperature$ = 0.85, HSL and CNT lead to similar predictions,
deviating from simulation results by two orders of magnitude.
At $\temperature$ = 0.95, a nucleation rate
of $\ln\nuclrate$ = -16.08 was obtained
for $\supersat$ = 1.146 (using $\threshold$ > 3 $\crit{\nuclsize}_\textrm{SPC}$)
where CNT predicts $\ln\nuclrateCNT$ = -19.99, cf.\ Tab.\ \ref{tabJ},
as opposed to $\ln\nuclrateHale$ = -24.27. 
Thus, HSL breaks down at high temperatures for the \LJTS{} fluid.
Present results generally agree with nucleation rates obtained by
\NVT{} simulation at temperatures between 0.65 and 0.95, as can be
seen by comparison with the SPC modification
that was correlated to data from canonical ensemble MD simulation \cite{HVH08}.

Figure \ref{szilVII} shows how the choice of $\threshold$ affects the
\nucleus{} temperature. The largest \nuclei{} allowed to remain in the system
have a highly elevated temperature and the amount of \nucleus{} overheating
can be explained by considering the boundary condition that
McDonald's \daemon{} imposes on size fluctuations. Only \nuclei{} that do not fluctuate
to sizes above $\threshold$ remain in the system for a significant time.
Almost all \nuclei{} with $\nuclsize \approx \threshold$
approach the point where overheating due to the enthalpy of
vaporization released during condensation countervails the
supercooling of the \vapour{}.
Note that this effect is much stronger than the overheating $\crit{\overheating}$ of the
critical \nucleus{} according to CNT due to nucleation kinetics \cite{FRLP66}
\begin{equation}
   \crit{\overheating}
      = \frac{2\zeldovich\kboltz\temperature^2}{\deltahvap},
\end{equation}
where $\zeldovich$ is the Z\v{e}l'dovi\v{c} factor and $\deltahvap$ is the enthalpy
of vaporization, evaluating to $\crit{\overheating}_\textrm{CNT}$ = 0.00608 in the present case.

With a threshold far below the critical size, the
intervention rate of McDonald's \daemon{} is several orders of magnitude
higher than the steady-state nucleation rate, cf.\ Tab.\ \ref{tabTheta} and Fig.\ \ref{szilV}.
In agreement with
Eq.\ (\ref{eqn:intervention}), $\nuclrateof{\threshold}$ reaches a plateau
for $\threshold > \crit{\nuclsize}_\textrm{SPC}$.
In particular, the approximation $\nuclrate \approx \nuclrateof{\threshold}$ is
valid for all values shown in Tab.\ \ref{tabJ} and Fig.\ \ref{szilIV}.
As Tab.\ \ref{tabTheta} also shows, the density and the pressure of the supersaturated
\vapour{} have very good convergence properties with respect to the intervention
threshold size and can already be accurately obtained at a high accuracy for
$\threshold$ values near the critical size.

\section{Conclusion}
GCMD with McDonald's \daemon{} was established as a method for steady-state simulation of
nucleation processes. The main purpose of the new method consists in directly simulating a
metastable state that undergoes a phase transition at a high rate without being
limited to sampling only the short timespan until nucleation occurs.

By implication, growth or decay processes of very large \nuclei{} are
not covered. These hava to be considered using the cutoff correction given by
Eq.\ (\ref{eqn:intervention}) unless the intervention threshold size is significantly
larger than $\crit{\nuclsize}$.
Due to an intervention scheme based on the
single order parameter $\nuclsize$, other relevant order parameters such as
shape or temperature of the \nuclei{} can experience a perturbation
for a \nucleus{} size similar to $\threshold$. It was shown for the
\nucleus{} temperature that this only concerns the largest \nuclei{} in the
system and that the range of \nucleus{} sizes
unaffected by intervention based overheating can be extended
arbitrarily if a sufficiently high value of $\threshold$ is chosen.

The intervention rate necessarily approaches the nucleation
rate for increasing values of the intervention threshold size. The
dependence of $\nuclrateof{\threshold}$ on $\threshold$ is
already accurately described for $\threshold > \crit\nuclsize\slash{}2$ by modeling the
\nucleus{} size evolution as a one-dimensional random walk without
taking any other order parameter into account.

For \vapour{} to liquid nucleation of the \LJTS{} fluid,
a series of simulations was conducted over a wide range of temperatures.
Good agreement with canonical ensemble MD simulation results was reached.
It was confirmed that CNT overstates the free energy of \nucleus{} formation and
underpredicts the nucleation rate.
HSL accurately describes nucleation near the triple point temperature;
at high temperatures, however, significant deviations are present.

\bigskip

\noindent \textbf{Acknowledgment.} 
   The authors would like to thank Martin Bernreuther (Stuttgart), Guram Chkonia (Cologne),
   Hans Hasse (Kaiserslautern), Svetlana Miroshnichenko (Paderborn), Srikanth Sastry (Bangalore), Chantal Valeriani (Edinburgh),
   and Jan Wedekind (Barcelona) for openly discussing methodological issues
   as well as Deutsche Forschungsgemeinschaft (DFG) for funding
   the collaborative research center (SFB) 716 at Universit\"at Stuttgart.
   The presented research
   was conducted under the auspices of the Boltzmann-Zuse Socie\-ty of
   Computational Molecular Engineering (BZS), and the simulations were
   performed on the HP XC4000 supercomputer at
   the Steinbuch Centre for Computing, Karlsruhe, under the grant LAMO.


\newpage

\begin{table}[t]
\caption{}
\label{tabJ}
\end{table}
\noindent
Average number of particles and intervention rate of McDonald's \daemon{}
during GCMD simulation as well as the nucleation rate approximated
by $\nuclrate \approx \probinftyof{\threshold}(\textrm{CNT}) \nuclrateof{\threshold}$, cf.\
Eqs.\ (\ref{eqn:committor}) and (\ref{eqn:intervention}), in dependence of
simulation conditions, i.e., temperature (in units of $\LJeps\slash\kboltz$),
supersaturation, intervention threshold size (in particles), and system volume
(in units of $\LJsig^3$), compared to theoretical predictions for the nucleation rate
according to CNT and the SPC modification; all logarithms are given with respect
to the reduced rates, normalized by $\mathnormal{(\mass\slash\LJeps)^{1\slash{}2}\LJsig^{-4}}$.
Note that the intervention threshold size is sufficiently
larger than the critical size in all cases.

\bigskip

\begin{center}
\begin{tabular}{cccc||ccc|cccc}
\hline\hline
   $\temperature$ & $\supersat$ & $\threshold$ & $\volume$
                  & $\absnum$ & $\ln\nuclrateof{\threshold}$ & $\ln\nuclrate$ 
                  & $\ln\nuclrate_\textrm{CNT}$ & $\ln\nuclrate_\textrm{SPC}$ & $\crit{\nuclsize}_\textrm{CNT}$ & $\crit{\nuclsize}_\textrm{SPC}$ \\ \hline 
   0.65 & 3.500 & \phantom{0}66 & 1.38 $\times$ $10^{7}$
        & \phantom{0\,}261\,000 & -21.12 & -21.12
        & -25.80 & -20.83 & \phantom{0}30 & \phantom{0}27 \\
        & 3.800 & \phantom{0}45 & 2.40 $\times$ $10^{7}$
        & \phantom{0\,}529\,000 & -18.42 & -18.42
        & -23.44 & -18.86 & \phantom{0}24 & \phantom{0}21 \\
        & 4.100 & \phantom{0}36 & 2.03 $\times$ $10^{7}$
        & \phantom{0\,}517\,000 & -16.29 & -16.33
        & -21.60 & -17.44 & \phantom{0}21 & \phantom{0}17 \\
        & 4.400 & \phantom{0}30 & 1.54 $\times$ $10^{7}$
        & \phantom{0\,}487\,000 & -13.79 & -13.80
        & -19.96 & -16.20 & \phantom{0}18 & \phantom{0}14 \\ \hline
   0.7\phantom{0} & 2.496 & \phantom{0}74 & 2.16 $\times$ $10^{7}$
        & \phantom{0\,}518\,000 & -22.08 & -22.08
        & -26.40 & -21.33 & \phantom{0}41 & \phantom{0}39 \\
        & 2.616 & \phantom{0}75 & 3.95 $\times$ $10^{7}$
        & 1\,040\,000 & -20.63 & -20.63
        & -24.50 & -19.70 & \phantom{0}36 & \phantom{0}32 \\
        & 2.692 & \phantom{0}72 & 1.85 $\times$ $10^{7}$
        & \phantom{0\,}518\,000 & -19.04 & -19.04
        & -23.48 & -18.83 & \phantom{0}33 & \phantom{0}29 \\
        & 2.774 & \phantom{0}60 & 3.51 $\times$ $10^{7}$
        & 1\,030\,000 & -18.24 & -18.24
        & -22.49 & -18.06 & \phantom{0}30 & \phantom{0}26 \\
        & 2.866 & \phantom{0}51 & 2.02 $\times$ $10^{6}$
        & \phantom{0\,0}63\,800 & -16.77 & -16.77
        & -21.49 & -17.30 & \phantom{0}27 & \phantom{0}23 \\
        & 2.959 & \phantom{0}45 & 1.50 $\times$ $10^{7}$
        & \phantom{0\,}513\,000 & -15.65 & -15.65
        & -20.59 & -16.62 & \phantom{0}25 & \phantom{0}21 \\ \hline
   0.85 & 1.426 & 219 & 1.68 $\times$ $10^{7}$
        & 1\,040\,000 & -19.62 & -19.62
        & -22.28 & -18.66 & \phantom{0}80 & \phantom{0}80 \\
        & 1.440 & 198 & 1.62 $\times$ $10^{7}$
        & 1\,030\,000 & -17.89 & -17.89
        & -21.51 & -17.98 & \phantom{0}74 & \phantom{0}72 \\
        & 1.461 & 186 & 1.87 $\times$ $10^{6}$
        & \phantom{0\,}125\,000 & -16.55 & -16.55
        & -20.48 & -17.10 & \phantom{0}66 & \phantom{0}62 \\
        & 1.483 & 144 & 6.24 $\times$ $10^{6}$
        & \phantom{0\,}431\,000 & -15.54 & -15.54
        & -19.41 & -16.20 & \phantom{0}58 & \phantom{0}54 \\ \hline
   0.9\phantom{0} & 1.240 & 209 & 3.45 $\times$ $10^{6}$
        & \phantom{0\,}256\,000 & -21.33 & -21.35
        & -23.24 & -20.33 & 137 & 149 \\
        & 1.260 & 162 & 3.23 $\times$ $10^{6}$
        & \phantom{0\,}255\,000 & -18.21 & -18.26
        & -21.29 & -18.37 & 110 & 116 \\
        & 1.280 & 127 & 2.98 $\times$ $10^{6}$
        & \phantom{0\,}247\,000 & -17.01 & -17.10
        & -19.72 & -16.91 & \phantom{0}90 & \phantom{0}91 \\ \hline
   0.95 & 1.146 & 564 & 4.88 $\times$ $10^{6}$
        & \phantom{0\,}516\,000 & -16.08 & -16.08
        & -19.99 & -17.89 & 156 & 175 \\
\hline\hline
\end{tabular}
\end{center}

\newpage

\begin{table}[t]
\caption{}
\label{tabTheta}
\end{table}
\noindent
Pressure supersaturation $\pressure\slash\sat\pressure(\temperature)$
and density supersaturation $\density\slash\sat\density(\temperature)$
as well as the intervention rate of
McDonald's \daemon{} in dependence of simulation conditions along with
the long-term growth probability $\probinftyof{\threshold}$ of a \nucleus{} containing
$\threshold$ particles, cf.\ Eq.\ (\ref{eqn:committor}),
according to CNT.

\bigskip

\begin{center}
\begin{tabular}{cccc||ccc|c}
\hline\hline
   $\temperature$ & $\supersat$ & $\threshold$ & $\volume$
                  & $\pressure\slash\sat\pressure(\temperature)$ & $\density\slash\sat\density(\temperature)$ & $\ln\nuclrateof{\threshold}$
                  & $\probinftyof{\threshold}(\textrm{CNT})$ \\ \hline 
   0.7 & 2.496 & \phantom{0}10 & 5.38 $\times$ $10^{6}$
       & 2.70 & 3.16 & -13.55
               & \phantom{$>$}3.98 $\times$ $10^{-7}$ \\
       & & \phantom{0}15 & 4.31 $\times$ $10^{7}$
       & 2.69 & 3.17 & -15.65
               & \phantom{$>$}4.61 $\times$ $10^{-5}$ \\
       & & \phantom{0}20 & 4.31 $\times$ $10^{7}$
       & 2.75 & 3.26 & -16.99
               & \phantom{$>$}1.25 $\times$ $10^{-3}$ \\
       & & \phantom{0}25 & 5.38 $\times$ $10^{6}$
       & 2.78 & 3.32 & -17.63
               & \phantom{$>$}0.01 \phantom{$\times$ $10^{-3}$} \\
       & & \phantom{0}30 & 2.16 $\times$ $10^{7}$
       & 2.77 & 3.31 & -19.20
               & \phantom{$>$}0.07 \phantom{$\times$ $10^{-3}$} \\
       & & \phantom{0}35 & 5.38 $\times$ $10^{6}$
       & 2.78 & 3.32 & -19.89
               & \phantom{$>$}0.20 \phantom{$\times$ $10^{-3}$} \\
       & & \phantom{0}48 & 4.31 $\times$ $10^{7}$
       & 2.78 & 3.33 & -21.74
               & \phantom{$>$}0.77 \phantom{$\times$ $10^{-3}$} \\
       & & \phantom{0}56 & 2.16 $\times$ $10^{7}$
       & 2.78 & 3.32 & -21.18
               & \phantom{$>$}0.95 \phantom{$\times$ $10^{-3}$} \\
       & & \phantom{0}65 & 4.31 $\times$ $10^{7}$
       & 2.78 & 3.32 & -21.90
               & $>$0.99 \phantom{$\times$ $10^{-3}$} \\
       & & \phantom{0}74 & 2.16 $\times$ $10^{7}$
       & 2.77 & 3.32 & -22.08
               & $>$0.99 \phantom{$\times$ $10^{-3}$} \\ \hline
   0.9 & 1.240 & \phantom{0}89 & 3.45 $\times$ $10^{6}$
       & 1.33 & 1.67 & -18.87
       & \phantom{$>$}0.04 \phantom{$\times$ $10^{-3}$} \\
       & & 149 & 3.45 $\times$ $10^{6}$
       & 1.34 & 1.69 & -19.80
       & \phantom{$>$}0.62 \phantom{$\times$ $10^{-3}$} \\
       & & 209 & 3.45 $\times$ $10^{6}$
       & 1.34 & 1.68 & -21.33
       & \phantom{$>$}0.98 \phantom{$\times$ $10^{-3}$} \\ \hline
   0.9 & 1.260 & \phantom{0}70 & 3.23 $\times$ $10^{6}$
       & 1.36 & 1.74 & -16.64
       & \phantom{$>$}0.04 \phantom{$\times$ $10^{-3}$} \\
       & & 116 & 3.23 $\times$ $10^{6}$
       & 1.37 & 1.78 & -17.82
       & \phantom{$>$}0.55 \phantom{$\times$ $10^{-3}$} \\
       & & 162 & 3.23 $\times$ $10^{6}$
       & 1.37 & 1.79 & -18.21
       & \phantom{$>$}0.95 \phantom{$\times$ $10^{-3}$} \\ \hline
   0.9 & 1.280 & \phantom{0}55 & 2.98 $\times$ $10^{6}$
       & 1.37 & 1.77 & -15.69
       & \phantom{$>$}0.04 \phantom{$\times$ $10^{-3}$} \\
       & & \phantom{0}91 & 2.98 $\times$ $10^{6}$
       & 1.39 & 1.87 & -16.08
       & \phantom{$>$}0.47 \phantom{$\times$ $10^{-3}$} \\
       & & 127 & 2.98 $\times$ $10^{6}$
       & 1.39 & 1.88 & -17.01
       & \phantom{$>$}0.91 \phantom{$\times$ $10^{-3}$} \\
\hline\hline
\end{tabular}
\end{center}

\newpage

\begin{figure}[h]
\centerline{
   \includegraphics[width=8.5cm]{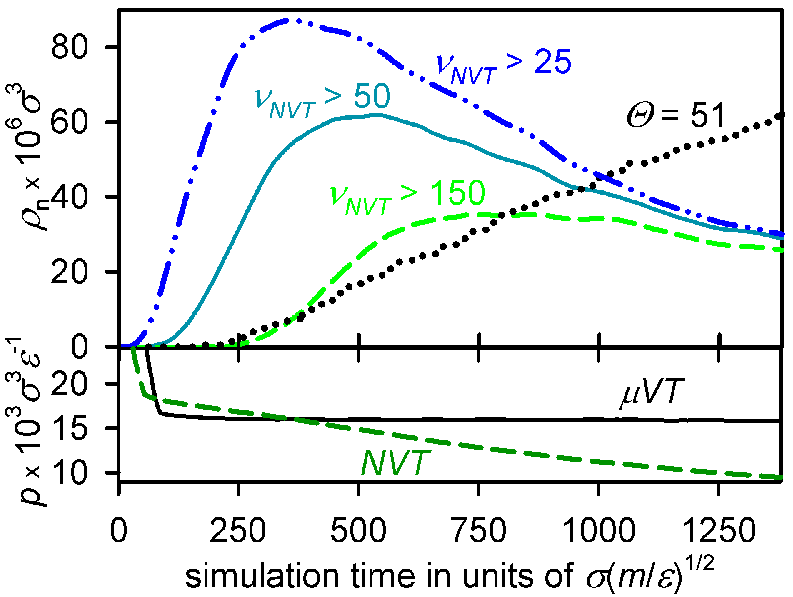}
}
\caption{}
\label{szilII}
\end{figure}

\begin{figure}[h]
\centerline{
   \includegraphics[width=8.5cm]{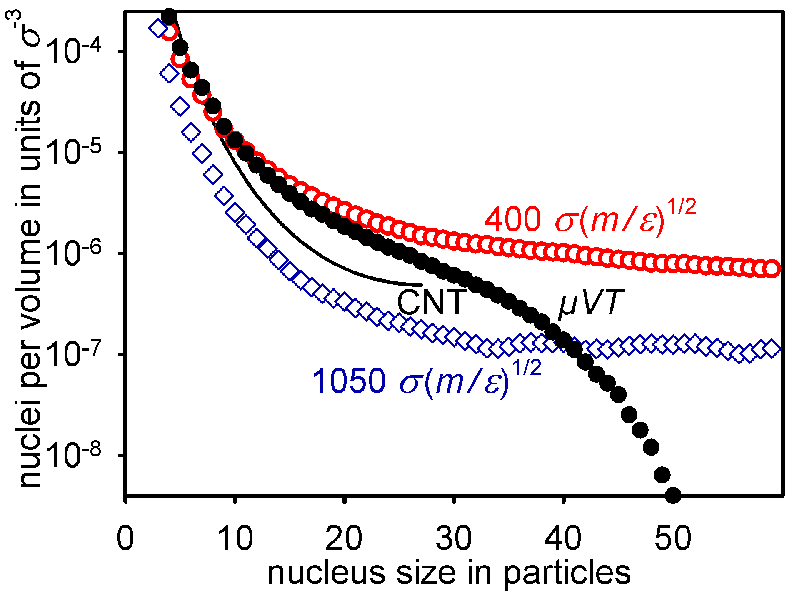}
}
\caption{}
\label{szilVI}
\end{figure}

\begin{figure}[h]
\centerline{
   \includegraphics[width=8.5cm]{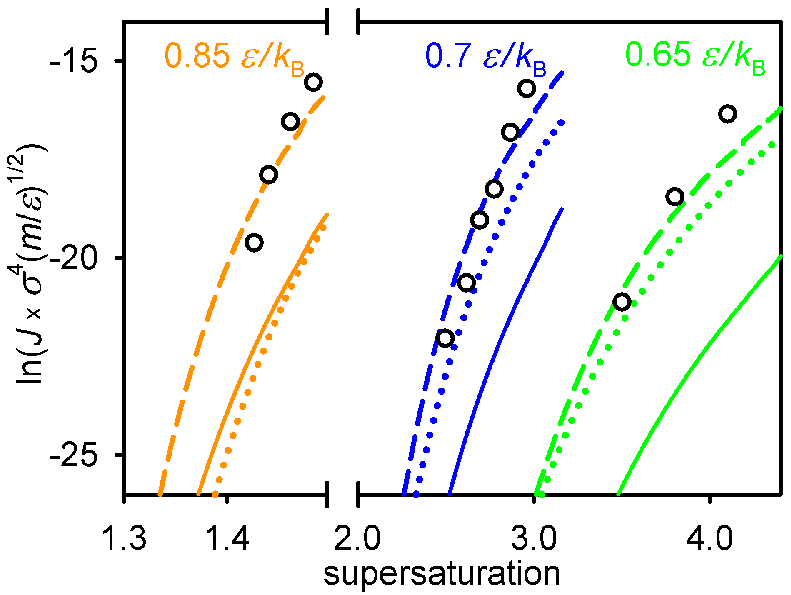}
}
\caption{}
\label{szilIV}
\end{figure}

\begin{figure}[h]
\centerline{
   \includegraphics[width=8.5cm]{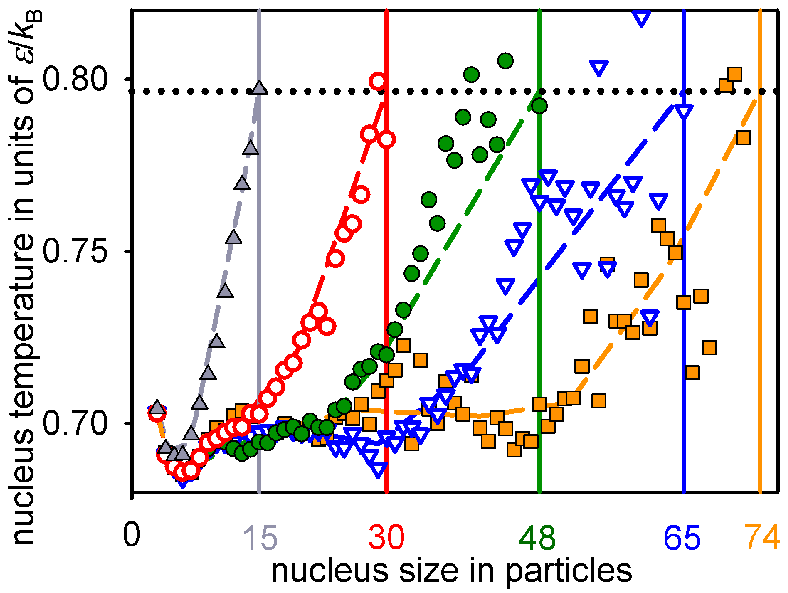}
}
\caption{}
\label{szilVII}
\end{figure}

\begin{figure}[h]
\centerline{
   \includegraphics[width=8.5cm]{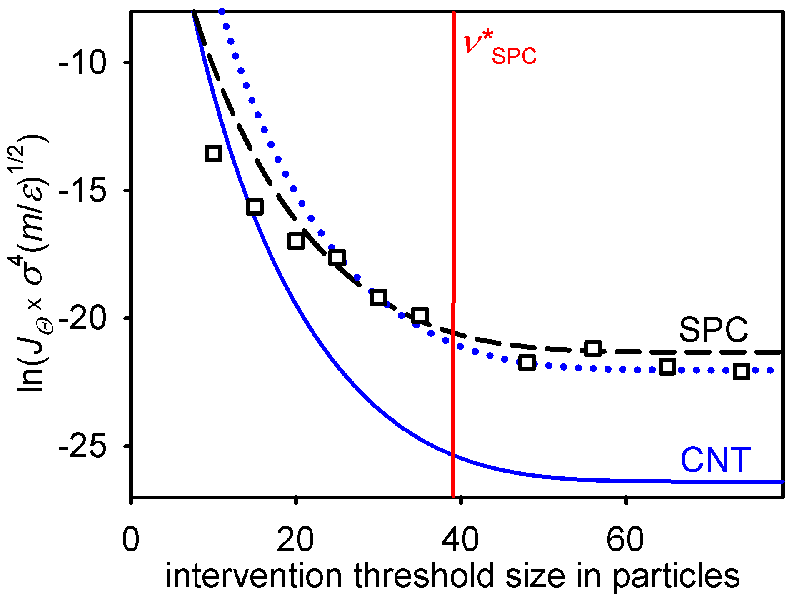}
}
\caption{}
\label{szilV}
\end{figure}

\begin{description}
   \item[Figure \ref{szilII}]
      Top: number per unit volume $\nucdensity$ of \nuclei{} containing $\nuclsize$ $>$
           25 $\textnormal{($\cdot$ $\cdot$ --)}$, 50 (---),
           and $\textnormal{150 (-- --)}$ particles in a $NVT$ simulation at
           $\temperature$ = 0.7 and $\density$ = 0.004044
           (in units of $\LJsig^{-3}$) as well as
           the aggregated number of McDonald's \daemon{} interventions per unit volume
           in a GCMD simulation with $\temperature$ = 0.7, $\supersat$ = 2.8658,
           and $\threshold$ = 51 ($\cdot$ $\cdot$ $\cdot$) over simulation time;
      bottom: pressure over simulation time for the $\NVT$
              simulation $\textnormal{(-- --)}$
              and the GCMD simulation (---).
   \item[Figure \ref{szilVI}]
      \Nucleus{} number per unit volume $\nucdensity$ over nucleus size $\nuclsize$
      from \NVT{} simulation at $\temperature$ = 0.7 and
      $\density$ = 0.004044, with sampling intervals of 320 $\leq$ $\timea$ $\leq$ 480 ($\circ$) and
      970 $\leq$ $\timea$ $\leq$ 1130 ($\diamond$) after
      simulation onset, and from GCMD simulation with
      $\temperature$ = 0.7, $\supersat$ = 2.8658, and $\threshold$ = 51 ($\bullet$)
      in comparison with a prediction for the same conditions based on CNT (---).
   \item[Figure \ref{szilIV}]
      Nucleation rate logarithm $\ln\nuclrate$
      over supersaturation $\supersat$ at $\temperature$ =
      0.65, 0.7, and 0.85 according to CNT (---),
      the SPC modification (-- --) as well as HSL $\textnormal{($\cdot$ $\cdot$ $\cdot$)}$
      compared to present GCMD simulation results ($\circ$).
   \item[Figure \ref{szilVII}]
      \Nucleus{} temperature over \nucleus{} size from
      GCMD simulation at $\temperature$ = 0.7
      and $\supersat$ = 2.4958 for an
      intervention threshold size of $\threshold$ = 15 ($\blacktriangle$),
      30 ($\circ$), 48 ($\bullet$), 65 ($\nabla$), and 74
      particles ($\blacksquare$); dotted line: saturation temperature $\sat\temperature$ = 0.7965 of
      the \vapour{} at constant pressure $\pressure$ = 0.134, which
      corresponds to the chosen supersaturation; dashed lines: guide to the eye.
   \item[Figure \ref{szilV}]
      Intervention rate logarithm
      $\ln\nuclrateof{\threshold}$
      over intervention threshold size $\threshold$ of
      McDonald's \daemon{} during GCMD simulation at
      $\temperature = 0.7$ and $\supersat = 2.4958$ ($\square$)
      in comparison with
      predictions based on CNT (---) and the SPC modification (-- --);
      dotted line: CNT prediction shifted to the actual value
      of the nucleation rate; vertical line: critical size according
      to the SPC modification.
\end{description}

\newpage

\end{document}